\documentclass{aastex631}

\newcommand{\DelThe}[1]{\Delta\theta_\mathrm{#1}}

\graphicspath{{./}{figures/}}
\usepackage{graphicx}
\usepackage{outlines}
\usepackage{enumitem}
\usepackage{amsmath} 
\setenumerate[1]{label=\arabic*}
\setenumerate[2]{label*=.\arabic*}
\setenumerate[3]{label*=.\arabic*}
\setenumerate[4]{label*=.\arabic*}

\begin{document}

\title{Toward Spectrum Coexistence: First Demonstration of the Effectiveness of Boresight Avoidance between the NRAO Green Bank Telescope and Starlink Satellites}

\author[0000-0001-5122-9997]{Bang D. Nhan}
\affiliation{National Radio Astronomy Observatory (NRAO),
520 Edgemont Road,
Charlottesville, VA 22903 USA}

\author[0000-0002-0786-7307]{Christopher G. De Pree}
\affiliation{National Radio Astronomy Observatory (NRAO),
520 Edgemont Road,
Charlottesville, VA 22903 USA}

\author[0009-0005-2746-9145]{Matt Iverson}
\affiliation{Space Exploration Technologies Corporation (SpaceX),
Hawthorne, CA USA}

\author[0000-0002-0320-6181]{Brenne Gregory}
\affiliation{Green Bank Observatory (GBO),
Green Bank, WV USA}

\author{Daniel Dueri}
\affiliation{Space Exploration Technologies Corporation (SpaceX),
Hawthorne, CA USA}

\author[0000-0001-5844-8359]{Anthony Beasley}
\affiliation{National Radio Astronomy Observatory (NRAO),
520 Edgemont Road,
Charlottesville, VA 22903 USA}


\author[0009-0008-5489-2215]{Brian Schepis}
\affiliation{Space Exploration Technologies Corporation (SpaceX),
Hawthorne, CA USA}



\begin{abstract}
NRAO and SpaceX have been engaged in coordinated testing efforts since
Fall 2021, including conducting experiments on different interference
avoidance schemes for the Karl G. Jansky Very Large Array (VLA) in New
Mexico, and the Green Bank Telescope (GBT) inside the National Radio
Quiet Zone (NRQZ) in West Virginia. The Starlink system is capable of
avoiding direct illumination of telescope sites with their adaptive
tasking to place downlink beams far away. Nevertheless, even satellites
operating in this mode can potentially present strong signals into the
telescope's receiver system if they pass close to the telescope's main
beam at the boresight. For additional protection, Starlink
satellites can either momentarily redirect or completely disable their
downlink channels while they pass within some minimum angular
separation threshold from the telescope's boresight, methods that are referred
to as ``telescope boresight avoidance''. In two separate experiments
conducted since Fall 2023, NRAO and SpaceX arranged to have the
GBT observe a fixed RA/Dec position in the sky, chosen to have a large
number of close-to-boresight Starlink passages. Preliminary analysis from these two experiments illustrate the feasibility of these avoidance methods to significantly reduce, if not eliminate, the negative impact of close-to-boresight
satellite passages. Importantly, these experiments demonstrate the value of continuing cooperative efforts between NRAO and SpaceX, and expanding cooperation between the radio astronomy and satellite communities more generally.
\end{abstract}



\section{Introduction}
\label{sec:intro}
For decades, typical radio-frequency interference (RFI) issues faced by radio astronomy observatories either came from on-site or from emitters near the horizon. These spurious signals arose from nearby electronics and later from cellular telephone towers or other transmitters placed in the environment of observatories. For these reasons, observatories have typically been situated in remote locations, with some sites being protected by specially defined zones. The most well-known zone in the United States is the National Radio Quiet Zone (NRQZ), which was established in 1958 by FCC Docket 11745 and encompasses parts of West Virginia, Virginia, and Maryland. Meanwhile, some locations, such as the Karl G. Jansky Very Large Array (VLA) in New Mexico, have established coordination agreements with known local transmitters (e.g. White Sands Missile Range) without being surrounded by a defined protected zone. For a time, these arrangements provided adequate protection. In recent years, however, the growth of large satellite constellations has challenged this framework since no location is too remote for global satellite networks \citep[e.g.,][]{walker2020satcon1, boley2021satellite}. Satellite-borne signals have had significant impacts on radio astronomy at least since the advent of the Russian Global Navigation Satellite System \citep[e.g.,][]{combrinck1994coexisting}, which severely impacted observations the 1612 MHz Hydroxyl line.

The advent of satellite constellations like Starlink, operated by SpaceX, have added a significant complication to what had been the operating environment of remote radio astronomy observatories. Starlink satellites operate in Low Earth Orbit (LEO, $\sim$~160-1,600~km) and transmit downward from their orbits in order to provide broad-band internet to ground-based receivers and to communicate with gateway stations. This new environment will require additional safeguards along with, in particular, coordinated and cooperative efforts for both passive (e.g., radio astronomy observatories) and active (e.g., satellite operators) spectrum users to meet their differing goals. The work described in these experiments is a response to the call from NSF's Spectrum Innovation Initiative (SII) to engage in creative methods for spectrum coexistence.

\subsection{Starlink Operational Parameters}
\label{sec:satlink_spec}
SpaceX began launching production Starlink satellites in late 2019 and shipping Starlink User Terminals (UTs) at scale in 2021. Starlink uses $\sim$~500~MHz bandwidth (14.0-14.5~GHz) for their UT uplinks, and $\sim$~2~GHz bandwidth (10.7-12.75~GHz) for their UT downlinks. Uplink UT signals use one of eight 62.5-MHz wide channels in the uplink band, with all UTs in a given geological service region (or cell) sharing the same uplink channel. During normal operations, all Starlink UTs in a given cell will dynamically switch uplink and downlink channels. Uplink transmissions use 60~MHz of each 62.5~MHz channel. The UT emissions towards the horizon are limited to -72.76~dBW/Hz in 60~MHz, which corresponds to a maximum equivalent isotropically radiated power (EIRP) of $\sim$~3.2~W \citep{eccreport271}.

Downlinks are transmitted in one of eight 250-MHz wide channels. The highest EIRP density (-15.0~dBW/4~kHz for satellites in 570~km orbits) occurs at the maximum slant ($40.5^{\circ}$ from nadir). Starlink satellites adjust transmitting power using phased-array beams, accounting for spreading loss and transmitting antenna gain to achieve a target power flux density (PFD) at each UT of -146~dBW/m$^2$/4~kHz. A UT can transmit with a maximum short-term duty cycle of 33\%, although the long-term duty cycle is limited to 11\% over 30 minutes (this is a radiation hazard requirement). In practice, the duty cycle is typically of the order of 1\% or less, depending on network traffic. For reference, Starlink’s downlink 250-MHz wide channels, denoted as Channel 1-8, start at the following frequencies: 10.7, 10.95, 11.2, 11.45, 11.7, 11.95, 12.2, and 12.45~GHz.

\subsection{Precursor Experiments and Current Developments}
\label{sec:precursors}
Since Fall 2021, NRAO has conducted two types of tests with SpaceX: coordinated and uncoordinated. The former requires formal prearrangement at a given time (typically 48~hours ahead of the scheduled observation) and a pointing direction for our telescopes to intersect with a specific group of satellite passages, as informed by SpaceX. The latter does not involve input from SpaceX; NRAO schedules periodic observations on our telescopes at the same pointing position to assess the evolving impact of satellite's downlink signals during normal operation. 

The first coordinated experiment took place in Fall 2021 with the testing of a Starlink UT at several locations near the VLA in Socorro, NM \citep{depree2023memo222}. This experiment showed clearly that the major RFI issue posed by the UTs were not their uplink signals (14.0-14.5~GHz), but the downlink signals from satellite to ground. In an effort to monitor and evaluate the impact of these space to ground transmissions (10.7-12.7~GHz), NRAO worked to install $\sim$~60 Starlink UTs on the Alamo Navajo Indian Reservation (T'iistsoh) located about 25~miles northeast of the VLA and began monthly uncoordinated tests (still ongoing). Early results of this experiment indicate minor impacts of the Starlink transmissions \citep{depree2023memo223}.

\begin{figure*}[h!]
\centering
\includegraphics[width=0.45\columnwidth]{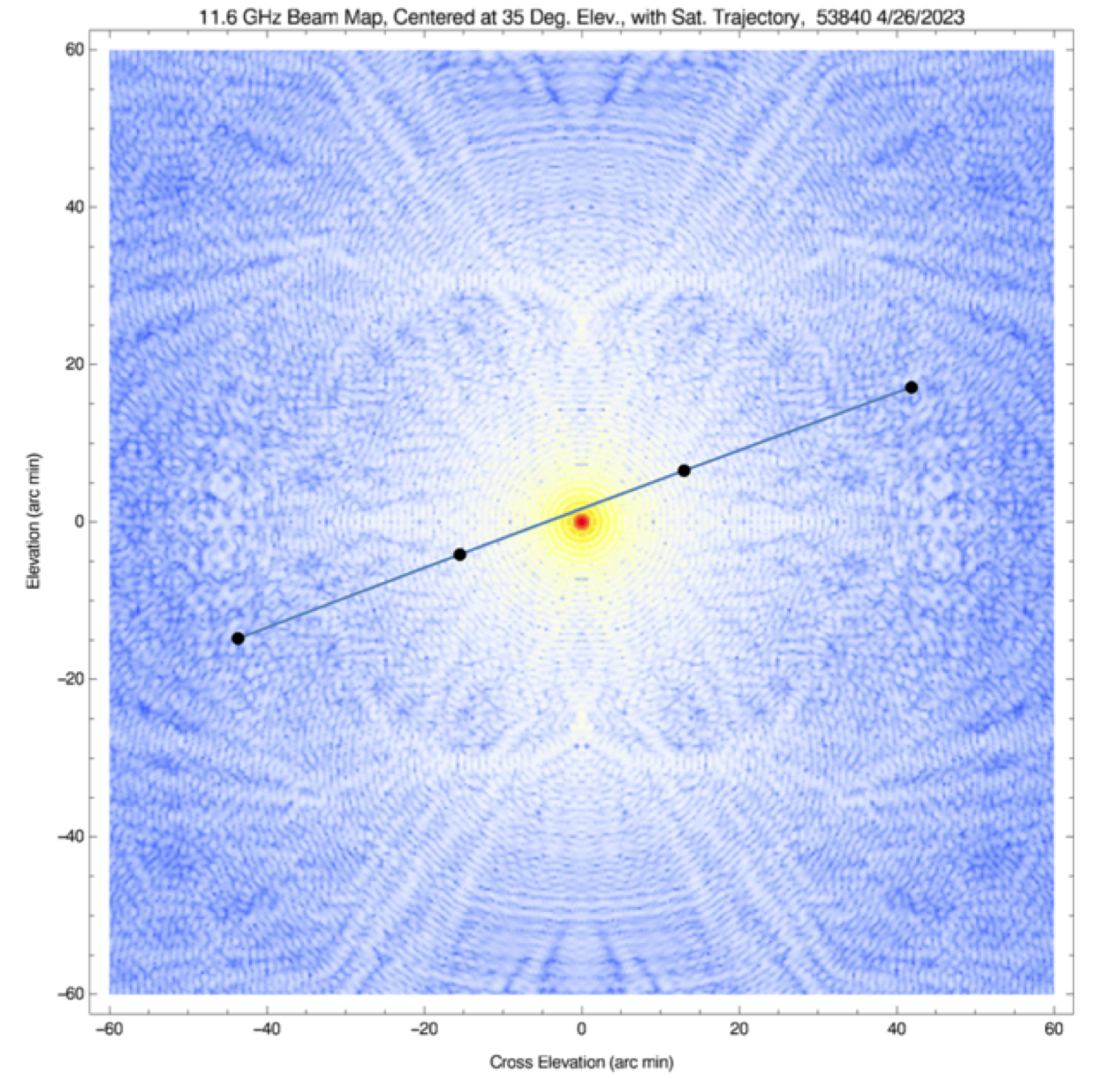}
\includegraphics[width=0.5\columnwidth]{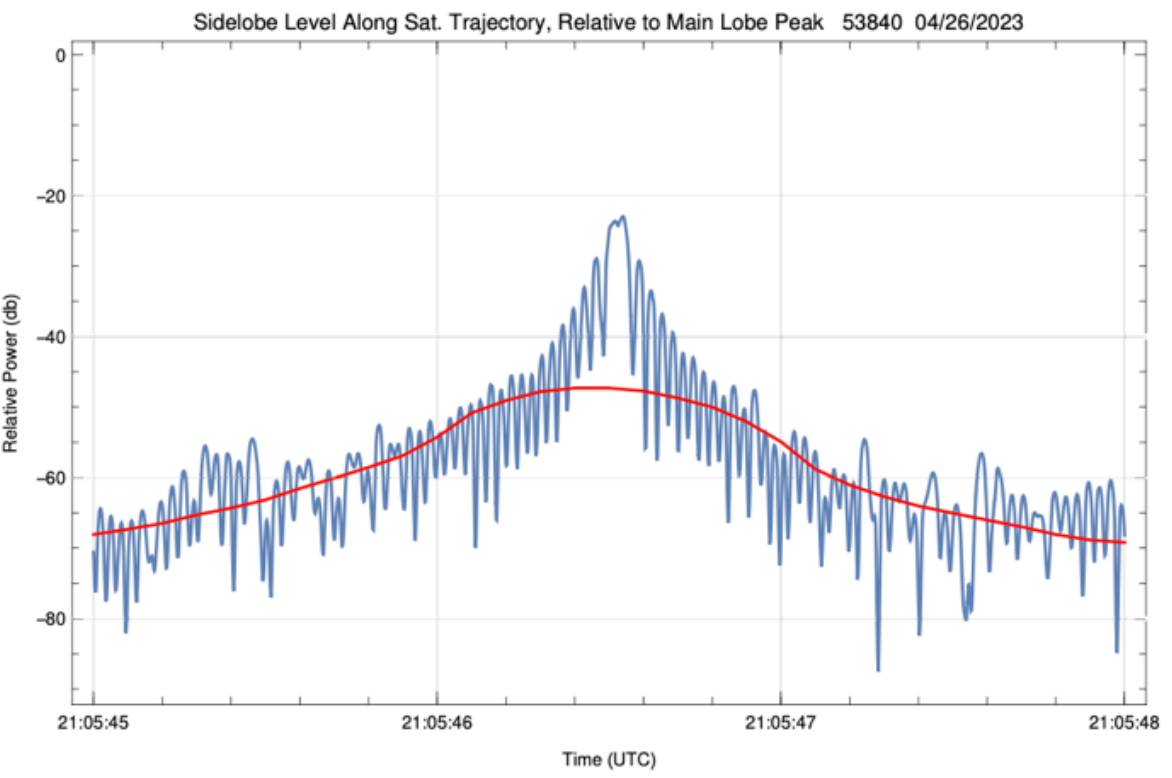}
\caption{(\textit{Left}) The simulated GBT beam map at 11.6~GHz, centered at the single dish antenna's elevation ($y$-axis) and cross-elevation ($x$-axis) angles which coincide with the azimuth and altitude angles of the satellite's trajectory, respectively, at the time of observations on April 26, 2023. The black dots indicate the orbital positions of a Starlink satellite at one second intervals, corresponding to the one-second integration time of the GBT observation, computed from public TLE. Note that this image covers a $2^{\circ}\times2^{\circ}$ square on the sky. (\textit{Right}) Profile plot of the simulated GBT's beam in Relative Power in decibels at 11.6~GHz (blue) and its one-second moving averaged
value (red curve). These figures were reproduced from \cite{depree2023memo154}.}
\label{fig:precursor_gbt_test}
\end{figure*}

Additionally, the impacts of direct site illumination was conducted through a series of coordinated tests in April and July of 2023. The Starlink's adaptive beam tasking placed their downlink beams to fully illuminate of the cells in the NRQZ. In the April test, the GBT detected a large signal while the telescope was pointed at an elevation of 25$^{\circ}$, and pointing to the North \citep{depree2023memo154}. Subsequent analysis of the Starlink satellites ephemeris, computed from publicly available Two-Line Elements (TLE), in the vicinity the GBT's boresight (Figure~\ref{fig:precursor_gbt_test}) verified that this large signal was the result of a satellite passing very close to the GBT's boresight at that moment. Since the Starlink satellites are at LEO altitudes, the downlink signal only intersects with the GBT's narrow main beam\footnote{Simulation from \cite{schwab2010memo271} provides GBT's FWMH(11.6 GHz) $\sim 0.86~\mathrm{arcmin} = 0.014^{\circ}$.} for a short amount of time, on the order of a few tenths of a second. In fact, this detection verified that a critical element to any coexistence scenario would be to account for and mitigate these close-to-boresight encounters by using any tasking capability available onboard, such as redirecting the downlink beams from the telescopes and possibly disabling downlink, when a satellite is within a certain angular separation from the telescope’s boresight during observation.

Recently, NRAO has been developing an autonomous telescope self-reporting system, the Operational Data Sharing \citep[ODS,][]{nhan2024ods} system, to provide near real-time telescope operational information queryable by satellite operators to be incorporated into their satellite tasking algorithms, including some of the telescope avoidance schemes. The ODS data will be frequently updated, ideally refresh every minute, to allow satellite systems, such as Starlink, to have awareness of the current pointing coordinates and observing frequencies of the VLA and GBT. A pilot testing program with the Starlink engineers to ingest the ODS data into their network algorithm is underway, but a detailed description of the program is outside the scope of this study. The boresight-avoidance experiments described in this study will play a key role in providing preliminary design requirements for the ODS system.

\section{Experimental Setup}
\label{sec:exp_setup}
\begin{figure*}[h!]
\centering
\includegraphics[width=0.45\columnwidth]{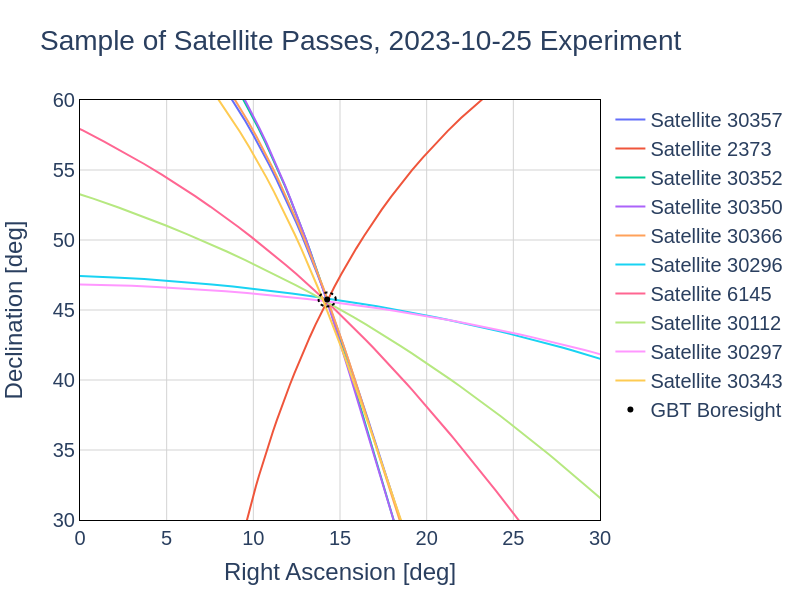}
\includegraphics[width=0.4\columnwidth]{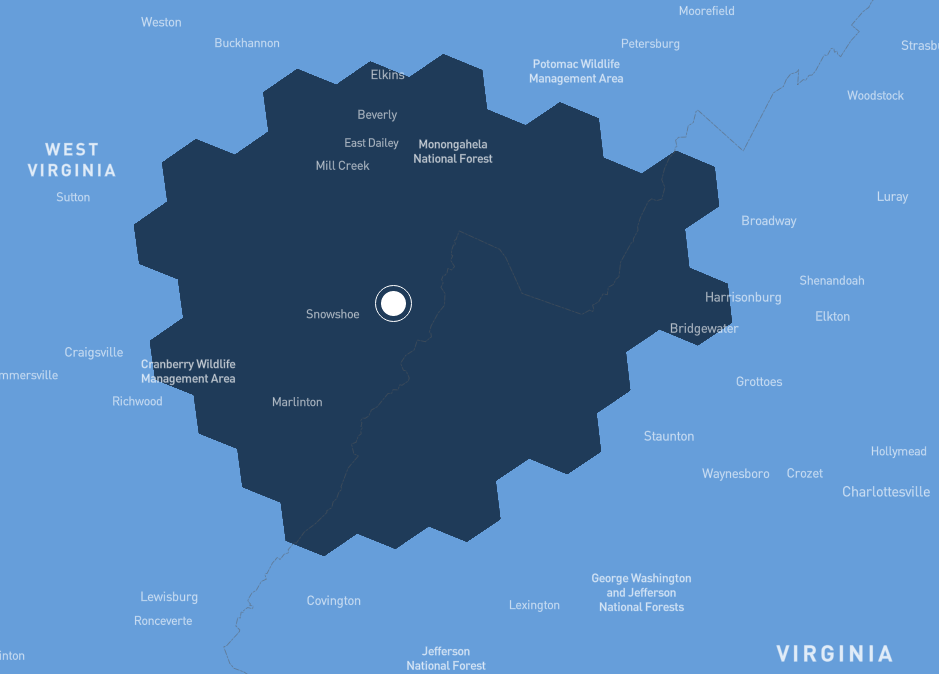}
\caption{(\textit{Left}) An example of the experimental setup for the Oct 2023 test, having the GBT track a fixed pointing at a sky region of (RA, Dec) = ($14.26^{\circ}, 45.75^{\circ}$), where a large number of satellite passages congregate within our observing window. The dashed circle indicates the 0.5$^{\circ}$ angular distance threshold from the GBT's boresight. (\textit{Right}) Service cells (dark blue), based on a hexagonal hierarchical geospatial indexing system, currently being voluntarily excluded by the Starlink system for fixed-address Starlink UTs within the NRQZ, as shown on the Starlink website.}
\label{fig:experiment_setup}
\end{figure*}

The coordinated experiments described in this study were carried out on October 25, 2023 (Experiment~\#1) and February 19, 2024 (Experiment~\#2). The observational parameters for the two experiments were identical, covering the downlink frequencies in four frequency bands and excluding the uplink bands. For both experiments,  the GBT was outfitted with an upgraded X-band receiver \citep[$\sim$~8-12~GHz,][]{morgan2024memo314} to make spectral measurements with the VEGAS\footnote{Versatile GBT Astronomical Spectrometer, \url{https://www.gb.nrao.edu/vegas/}} in four overlapping 1.5-GHz wide spectral windows centered at: 9.75, 10.5, 11.25, and 12~GHz, with a resolution bandwidth (RBW) of $\sim$91.6~kHz. These bands are respectively denoted as Spectral Bank (SpB) A, B, C, and D. As part of the coordination, the SpaceX engineering team provided a specified telescope pointing to track in equatorial coordinates (RA/Dec) over a given UTC time window to schedule an observation on the telescope beforehand. The telescope pointing was chosen to be a position that would result in a large number of close-to-boresight encounters using SpaceX's predicted satellite ephemeris (Figure~\ref{fig:experiment_setup}, center panel). The SpaceX team also provided a list of satellites with the estimated time when their angular separation from boresight ($\DelThe{bs}$) would be the smallest, based on their ephemeris. Satellite positions provided by SpaceX were consistent with those independently derived from the public TLE. 

In each experiment, the telescope observing script was executed by the authors manually in real time using the GBT's AstrID\footnote{Astronomer's Integrated Desktop, \url{https://www.gb.nrao.edu/softdev/}} and CLEO\footnote{Control Library for Engineers and Operators,\url{https://www.greenbankobservatory.org/~rmaddale/GBT/CLEOManual/index.html}} user interface software. Each observation consisted of two main steps. The first step was to conduct an auto-peak focus observing scan at a nearby source from the given pointing to correct for the pointing offset and the receiver feed's focus distance relative to GBT's secondary reflector. This procedure was followed by the main observation sequence with recorded data grouped into multiple scans, each accumulated ten-minutes worth of one-second long integrated spectra, for a total of 600 spectra per scan. 

One of the primary goals of this study is to assess potential impacts of Starlink satellites' downlink and their boresight avoidance capability for a representative GBT observation session. Although the VEGAS can sample spectra with much shorter time interval for high time resolution observation, most observations done on the GBT are commonly observed at one-second integration or longer for better signal-to-noise ratio (SNR). On rare occasions, such as when doing pulsar observation, that the observers would store spectra at sub-millisecond scale. We have planned to conduct follow-up observations at such a fine time scale to more accurately track the downlink signal from the fast-moving LEO satellites. Additionally, since obtaining the absolute flux calibration was not the priority in these tests, also due to the fast transition of the satellite passages over the telescope, we did not allocate time to observe a stronger astronomical flux density calibrator. 

\subsection{Experiment~\#1 - Boresight avoidance deactivated}
\label{sec:exp_1}
This observation started at around 22:09~UTC. The telescope took the first 17~minutes to slew to the prearranged pointing position at (RA, Dec) = (14.257$^{\circ}$, 45.745$^{\circ}$) = (00h57m1.6s, +45d44m43.00s). The auto-peak scan used the calibrator source 1824+5651 (1.15~Jy at 9.0~GHz) at 5.3$^{\circ}$ away from the pre-arranged pointing, to determine the focus correction of -9.59~mm and (azimuth, elevation) offset of (0.1660$^{\circ}$, 0.1885$^{\circ}$). This telescope pointing was provided by SpaceX to be a position that would result in 52 close-to-boresight encounters, ranging from $0.06^{\circ}$ to $2.96^{\circ}$ from the telescope's boresight. The main observation started at 22:25~UTC and concluded at 23:29~UTC.

During this experiment, the Starlink system (as it usually does) excluded the three cells located closest to the Green Bank Observatory (GBO), in order to avoid direct illumination of the GBT, but was available to serve mobile units (UTs for RVers and campers) using Channels 7 and 8 only within cells outside the three excluded ones. Also, the Starlink system excluded the illumination of any registered Starlink UTs in neighboring cells within the NRQZ listed as “not served” on the Starlink website\footnote{\url{https://www.starlink.com/map}} (Figure~\ref{fig:experiment_setup}, right panel). Starlink Generation 2 (Gen2) satellites equipped with downlink capability in Channels 1 and 2 are designated with a five-digit identification numbers starting with a ``3''. The Gen2 satellites were enabled to downlink in all eight Channels during this test, while all the Gen1 satellites were only capable of using Channels~3-8. The only major modification intended for the Experiment~\#1 to normal Starlink operations was the use of Channels~1 and 2 to serve locations greater than 180~km from the GBO site (approximately the distance between GBO and Washinton, DC). No boresight-avoidance scheme was implemented in this experiment.

\subsection{Experiment~\#2 - Boresight avoidance activated}
\label{sec:exp_2}
This observation started at around 20:08~UTC. The telescope took the first 17 minutes to slew to the pointing at (RA, Dec) = (90.1$^{\circ}$, 55.4$^{\circ}$) = (06h00m23.9s, +55d24m00.00s). The auto-peak scan used the calibrator source 0541+5312 (0.77 Jy at 9.0~GHz) located 3.5$^{\circ}$ away from the pre-arranged pointing, to determine the focus correction of -3.726~mm and (azimuth, elevation) offsets of (0.2911$^{\circ}$, -0.0081$^{\circ}$). This sky position was chosen so that during the test, there would be 49 close-to-boresight encounters, ranging from $0.17^{\circ}$ to $2.97^{\circ}$ from boresight. The main observation started at 20:25~UTC and concluded at 21:59~UTC.

Experiment~\#2 had the same parameters as described for Experiment~\#1, however, this time the Starlink network disabled downlink beams from satellites that were scheduled to pass within 0.5$^{\circ}$ of the GBT's boresight. This boresight angular separation threshold is experimental, and would be subject to change from satellite operator to operator due to their distinct onboard capabilities and service requirements. Before and after the period of this experiment, the Starlink network operated normally without any beam avoidance activated nor downlinking with Channels 1 and 2. 

\section{Results} 
\label{sec:results}

\subsection{Differential spectra} 
\label{sec:diff_spectra}
The spectral plots shown below were made by computing the difference between the raw spectra recorded at each one-second long integration and the median value of all the measured spectra. The raw (uncorrected) antenna temperature, $T_\mathrm{ant}(t,\nu)$, is first estimated as \citep{curtis2005memo237},  
\begin{equation}
    T_\mathrm{ant}(t,\nu) = 0.5T_\mathrm{cal}\frac{C_\mathrm{cal-ON} + C_\mathrm{cal-OFF}}{C_\mathrm{cal-ON} - C_\mathrm{cal-OFF}},
    \label{eq:count2Tant}
\end{equation}
where $T_\mathrm{cal}$ is the internal calibrator temperature, $C_\mathrm{cal-ON/OFF}$ are measured raw counts with internal noise diode calibration signal with a switching cycle of 1~Hz injected into the main RF signal path. Subsequently, the raw flux density, $S(t,\nu)$, in Jansky (Jy) were obtained by,
\begin{equation}
S(t,\nu) = \frac{T_\mathrm{ant}}{G},
\label{eq:flux_jy_convert}
\end{equation}
where the antenna gain $G = 2.84\eta_\mathrm{ap}$~[Kelvin/Jy] is computed by assuming the GBT's aperture efficiency, $\eta_\mathrm{ap} = 0.71$ across the entire band, and assuming no atmospheric correction is needed for the X-band. Since Equation~\eqref{eq:flux_jy_convert} assumes maximum gain of the GBT, this calculation of $S(t,\nu)$ could have underestimated the exact flux density of the satellite's downlink transmission. To remove any potential offset in individual spectra, we instead evaluated the differential flux density, which is computed as the difference between all the recorded spectra during the observation and their minimum value of the spectra over the observing time,
\begin{equation}
\Delta S(t,\nu) = S(t,\nu) - \mathrm{min}[S(t,\nu)],
\label{eq:delta_flux}
\end{equation}
where the minimum value of the spectra was chosen to ensure the least possible RFI is present in the reference spectrum.

\begin{figure*}[h!]
\centering
\includegraphics[width=1\columnwidth]{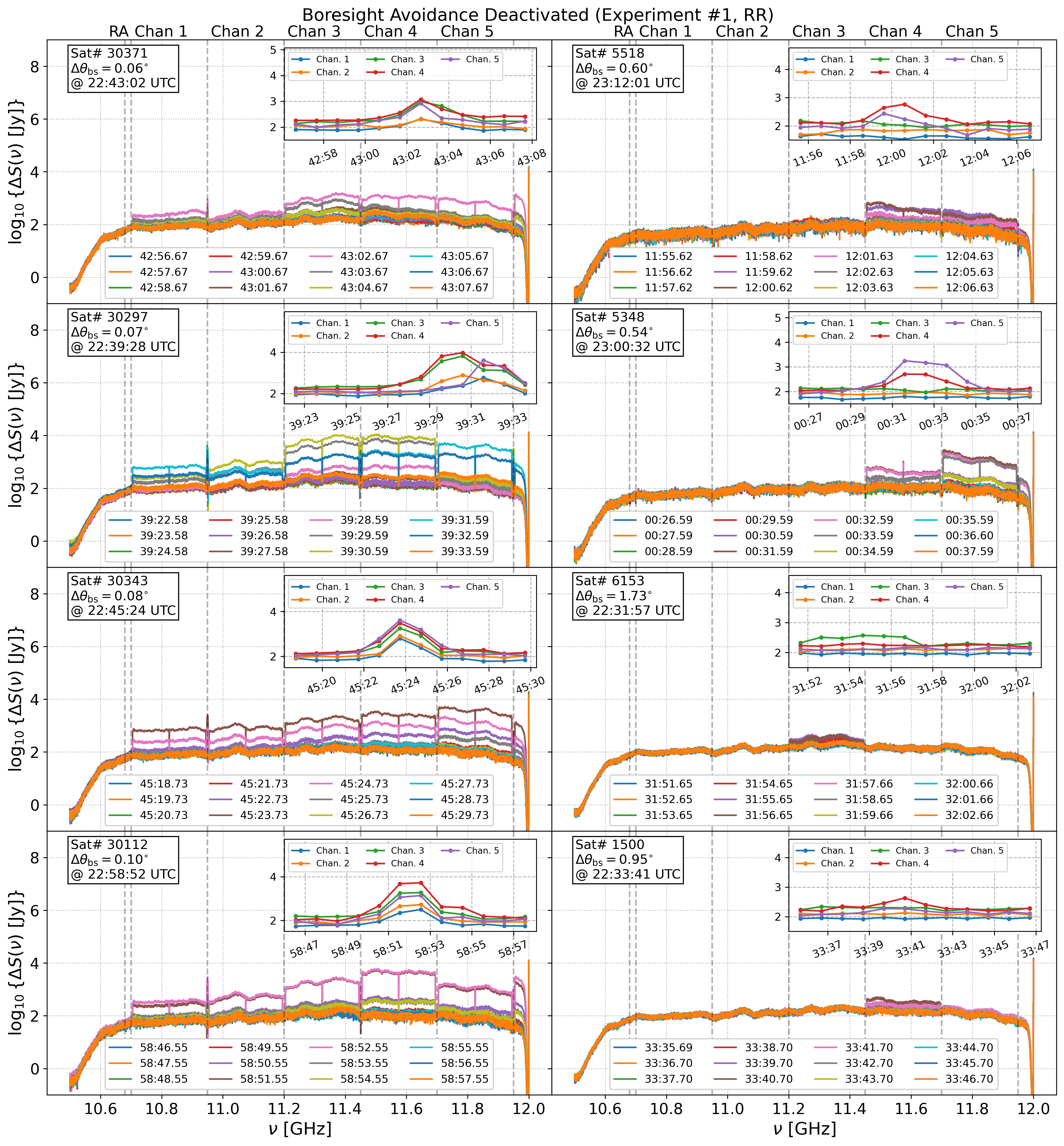}
\caption{Examples of the spectra from the right-handed circular polarization (RR) recorded in the SpB-C (10.6-12.0~GHz) during Experiment \#1, when boresight avoidance was not activated. (\textit{Left}) Observed spectra recorded with the RA-band, downlink Channel 1-5 with Starlink passages for $\DelThe{bs} \leq 0.5^{\circ}$. (\textit{Right}) Observed spectra recorded in the SpB-C with Starlink passages for $\DelThe{bs} > 0.5^{\circ}$. Satellite passages are shown with the satellite ID and boresight separation in degrees annotated in the legends. Note that it is simply a coincidence that the four satellite passages shown here are Gen2 satellites. We have observed Gen1 crossing the boresight with $\DelThe{bs} \leq 0.5^{\circ}$, similar to ones shown in Figure~\ref{fig:exp2_plots}. Additionally, each panel contains an inset showing the measured power (in logarithmic scale) at one of the passband frequencies within each downlink channels over twelve seconds, centered about the closest passage. In this experiment, the rise and fall of the power correlate with the satellite passages traversing the GBT's main beam, which will be traced out more accurately with higher time resolution in future experiments. Depending on the positioning of the downlink beams relative to the telescope, not all passages peak at the same time as when the satellites are closest to the telescope boresight.}
\label{fig:exp1_plots}
\end{figure*}

For illustration purposes, Figure~\ref{fig:exp1_plots} shows a subset of spectra recorded in the SpB-C centered at 11.25~GHz, covering the downlink Channels~1-5 and the protected radio astronomy (RA) band between 10.68-10.70~GHz. Examples of the GBT spectra recorded at instances when close-to-boresight Starlink passages with $\DelThe{bs} \leq 0.5^{\circ}$ (left panels) and $\DelThe{bs} > 0.5^{\circ}$ (right panels) are shown. Each panel shows eight spectra of one-second integration, centered at the time of the closest encounter in a given passage to show the varying signal strength as a satellite orbiting passes through the GBT's main beam. Note that strong narrow-band emissions are present at the transitions between channel pairs (Channels 1-2, 3-4 and 5-6) in some of the passages. The nature and a potential fix for this narrow-band emissions are discussed below.

\begin{figure*}[h!]
\centering
\includegraphics[width=1\columnwidth]{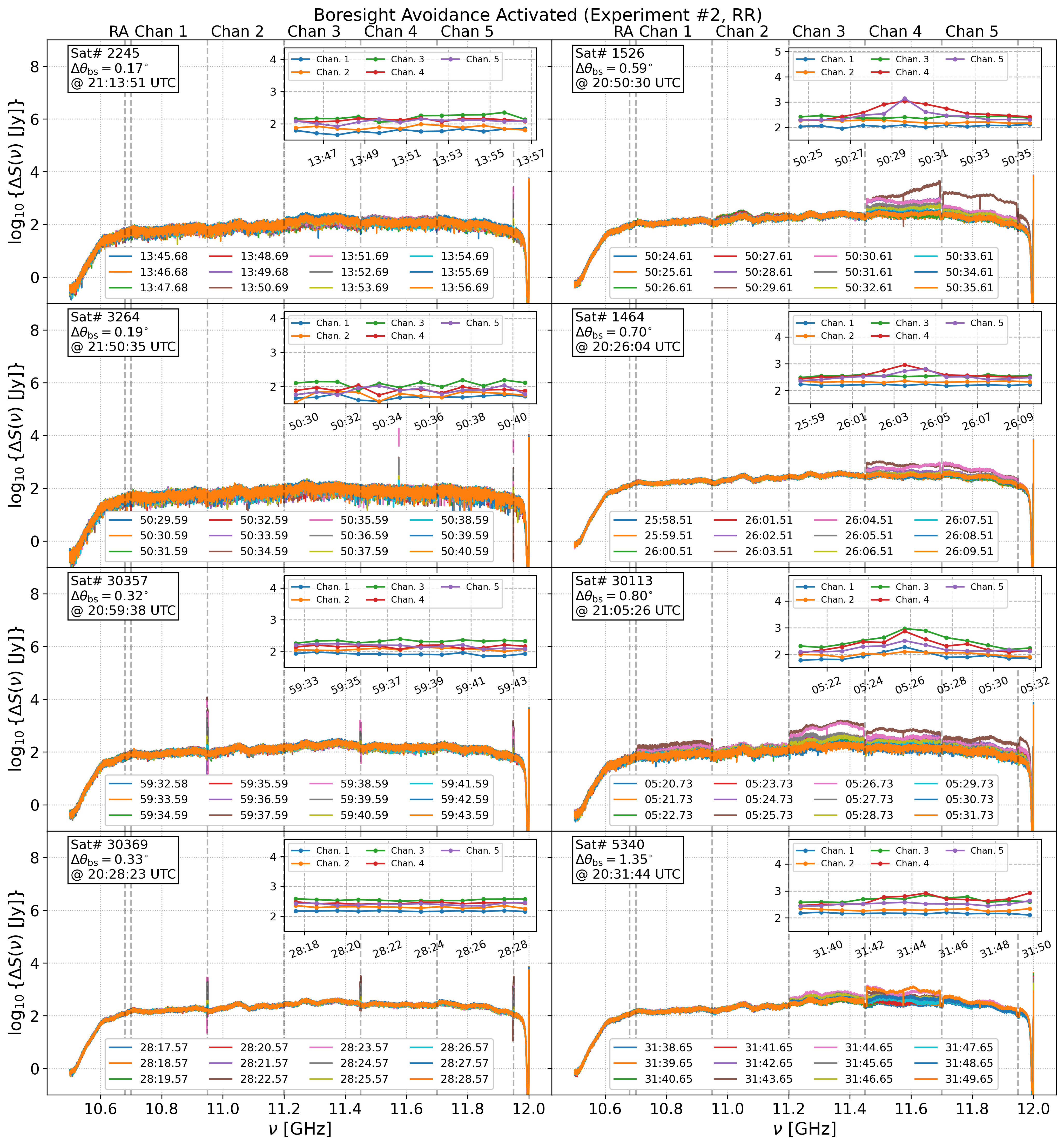}
\caption{Examples of the spectra from the right-handed circular polarization (RR) recorded in the SpB-C (10.6-12.0~GHz) during Experiment \#2, when the boresight avoidance was activated for Starlink passages with $\DelThe{bs} \leq 0.5^{\circ}$. (\textit{Left}) Observed spectra recorded with the RA-band, downlink Channel 1-5 with Starlink passages for $\DelThe{bs} \leq 0.5^{\circ}$. (\textit{Right}) Observed spectra recorded in SpB-C with Starlink passages for $\DelThe{bs} > 0.5^{\circ}$. Similarly, each panel contains an inset showing the measured power at one of the passband frequencies within each downlink channels over twelve seconds, centered about the closest passage. In this experiment, the rise and fall of the power are only observable for passages with the closest boresight separation outside the $0.5^{\circ}$ cutoff.}
\label{fig:exp2_plots}
\end{figure*}

In contrast to Figure~\ref{fig:exp1_plots}, Figure~\ref{fig:exp2_plots} shows the lack of apparent strong broad-band emission in Channels 1-5 for the spectra measured for close-to-boresight Starlink passages with $\DelThe{bs} \leq 0.5^{\circ}$. However, as in Experiment \#1, strong narrow-band emissions present at the transitions between downlink channels pairs remains even when the boresight avoidance was activated. These narrow-band emissions so far do not seem to pose a problem in our data since we do not observe any induced inter-modulation harmonic components or direct power leakage into the RA-band at  10.68-10.70~GHz. As part of the collaborative effort, after our reporting, the SpaceX team identified the source of these emission features and devised a software patch to suppress them, which will be validated by our follow-up observations. This example demonstrates the importance of coordination and collaboration between passive and active spectrum users, so both parties can resolve such issues in the early development and deployment phases.

\subsection{Signal-to-Noise Ratio Comparison} 
\label{sec:snr_compare}
To quantify the effectiveness of the boresight avoidance, beyond visual inspection of the spectra, two metrics are compared between the two experiments. Firstly, at an expected time for a close Starlink passage, we compute the relative SNR between the observed signal level in one of the eight downlink channels and the signal level of a clean channel far from the downlink channels. The clean channel is chosen to be between 9.60-9.80~GHz, observed in the SpB-A of the data. This clean channel is denoted as the reference (Ref) signal. Secondly, we assess the potential spectral leakage from the out-of-band emission (OOBE) into the 10.68-10.70~GHz RA-band when there is a nearby Starlink downlink in Channels 1-2 (10.70-11.20~GHz) by computing the SNR at the RA-band using the same clean reference signal at 9.6-9.8~GHz. 

\begin{figure*}[htb!]
\centering
\includegraphics[width=0.49\columnwidth]{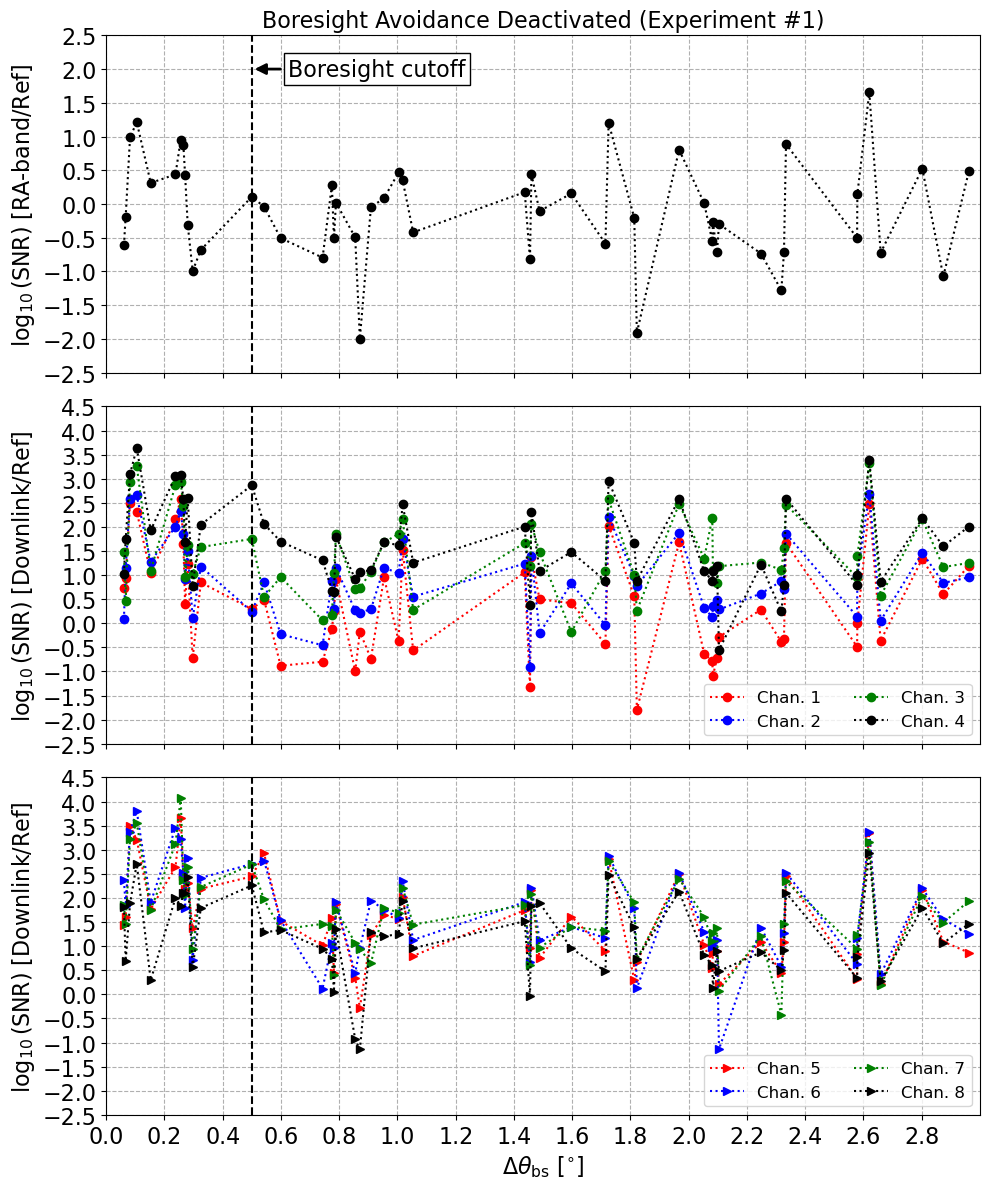}
\includegraphics[width=0.49\columnwidth]{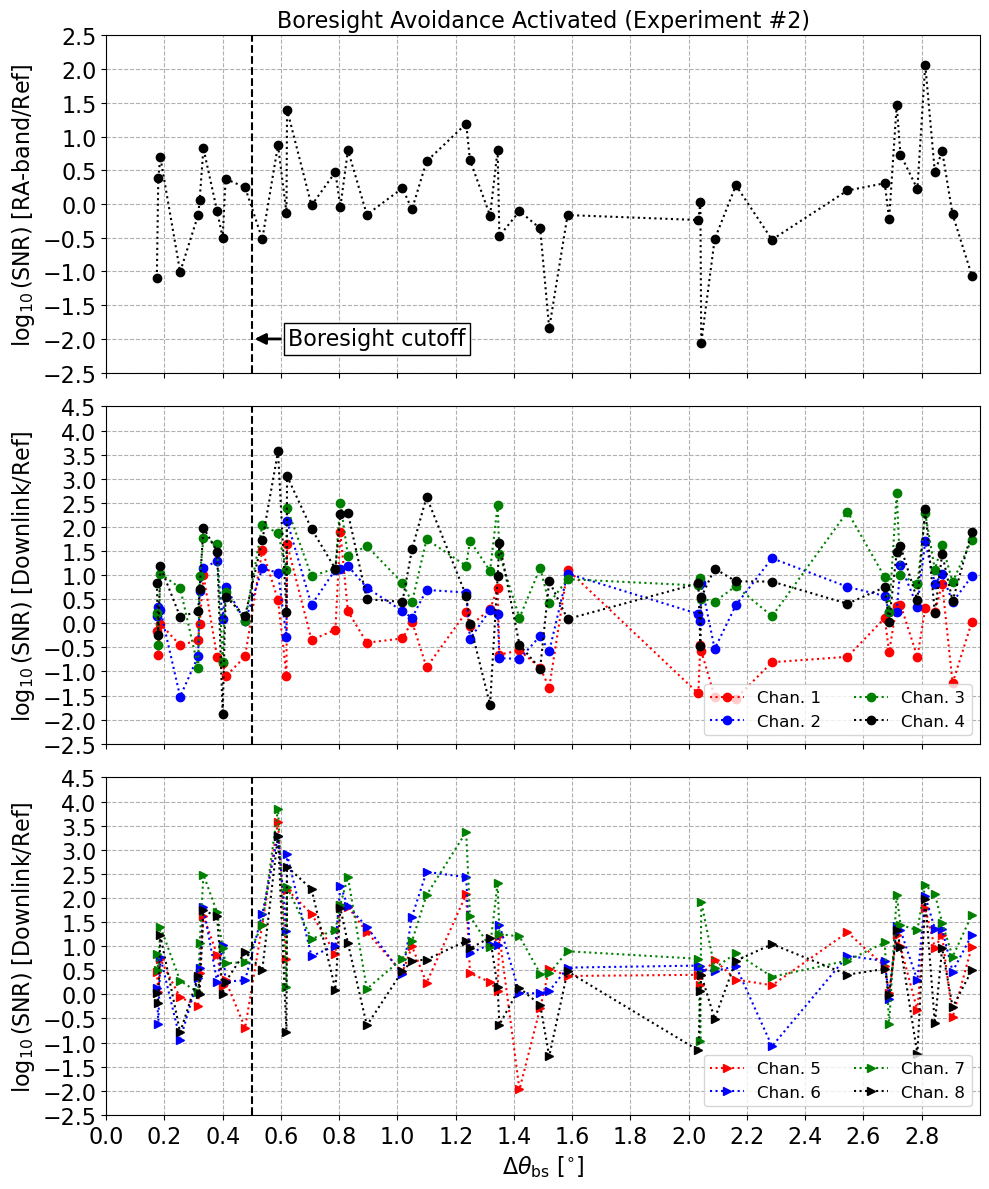}
\caption{Comparison of SNR (in logarithmic scale) between Experiment \#1 (\textit{Left}) and \#2 (\textit{Right}). SNR within the boresight separation threshold, $\DelThe{bs} = 0.5^{\circ}$ (dashed line), is distinctly lower in Experiment \#2 when the boresight-avoidance tasking is activated by the Starlink system. Each data point represents a unique satellite passage observed during the experiments, with a total of 52 (49) passages observed in Experiment \#1 (\#2). Note that the SNR for Experiment \#2 appears to be higher than the nominal SNR at $\DelThe{bs} = 0.59^{\circ}$ because a particular satellite passage (Sat \#1526 at 20:50:30 UTC, upper right panel of Figure~\ref{fig:exp2_plots}) was downlinking in Channel 4, 5, 6, 7, and 8.}
\label{fig:snr_compare_plot}
\end{figure*}

There are instances that downlink transmission from satellite systems other than Starlink present in our data. Hence, only spectra measured at the timestamps when the Starlink satellites were expected to be the closest to the GBT's pointing are considered. Since the downlink channel has a bandwidth of 250~MHz, the median level of the signal across a given downlink channel is used to compute the SNR. Namely, for a given downlink Channel~$j$, SNR$_j$ is defined as 
\begin{equation}
\mathrm{SNR}_j = \frac{\mathrm{med}[\Delta S_j(\nu)]}{\mathrm{med}[\Delta S_\mathrm{Ref}(\nu)]},
\label{eq:snr_def}
\end{equation}
where $S_\mathrm{Ref}$ is the flux density in the clean reference channels at 9.60-9.80~GHz.The SNR for the RA-band is computed in the same manner using the median value of the flux density across the 10.68-10.70~GHz band.

If boresight avoidance is activated by the Starlink system and working as expected, the observed SNR in the downlink channels, along with one in the RA-band, are close to unity (zero in logarithmic scale) since no transmission is made during the close-to-boresight passages. Our analysis shows that the boresight avoidance is functional and helps to reduce the RFI significantly. In particular, as shown in Figure~\ref{fig:snr_compare_plot}, for passages with $\DelThe{bs} \leq 0.5^{\circ}$, the SNR for the eight downlink channels in Experiment~\#1 (left panels) are at least two orders of magnitude higher than the ones in Experiment~\#2 (right panels) with boresight avoidance activated. 

\section{Discussion and Summary} 
\label{sec:discussion}
Besides avoiding direct site illumination, the primary method to protect a telescope from Starlink satellite transmissions is through adaptive beam tasking that places a satellite's downlink beams far away from the telescope site when the satellite is within a certain angular separation from the telescope's boresight during observation. For example, a satellite that passes within $2^{\circ}$ of boresight could be directed to steer its beams no closer than 180~km from a radio telescope. An additional protection level can be achieved by completely disabling downlink beams from satellites that pass within an even tighter cone of a telescope's boresight during observation. This operational mode would further reduce the chance of a telescope main beam being illuminated by any satellite's downlink beam, including its inner sidelobes. At the moment, these two mitigation methods are referred to, both separately and collectively, as the ``telescope boresight-avoidance'' method by the NRAO and SpaceX collaborators. A separate designation for each of these two methods will be adopted in the follow-up studies.

This study has, for the first time, demonstrated the feasibility of implementing the telescope boresight-avoidance method between a radio astronomy telescope and a LEO satellite operator through collaborative experiments between NRAO's Green Bank Telescope and SpaceX's Starlink constellation. This experiment was made possible by sharing the radio telescope's pointing position and frequency of observation with the LEO satellite operator, who was then able to use this data to mitigate interference into the telescope. The two experiments conducted at the GBT in October 2023 and February 2024 demonstrated:
\begin{itemize}
  \item When informed about a telescope's pointing direction and the frequency band being observed, the Starlink system is capable of disabling downlink beams for satellite passages close to telescope boresight. While this action is planned for the closest of boresight passages, we expect that refraining from placing beams near the radio telescope will suffice for most near-boresight passages of consequence.
  \item Briefly disabling satellite downlinks as a satellite passes close to boresight can significantly reduce the observed satellite emission in our data, indicated by statistically significant reductions in SNR by two orders of magnitude inside the $0.5^{\circ}$ radius. 
  \item For Starlink Gen2 passages using Channels 1 and 2, although the SNR levels of the RA-band between 10.68-10.7~GHz in both experiments are approximately unity, a closer inspection suggests a slight increase (about a factor of three) in signal level in Experiment \#1 for passages with $\DelThe{bs} \leq 0.5^{\circ}$ (Figure~\ref{fig:snr_compare_plot}, top left panel). This potential leakage is no longer an issue when boresight avoidance is in use for close passages (Figure~\ref{fig:snr_compare_plot}, top right panel). 
 
\end{itemize}

Although the boresight cutoff value of $\DelThe{bs} = 0.5^{\circ}$ was adopted in this study, it is subject to change as NRAO and SpaceX iterate on a final value for a future operational system. It is also worth noting that boresight avoidance is only in use for a small number of satellite passages that intersect telescope boresight (11 of 49 passages or 22.45\% in Experiment \#2). Adaptive beam tasking will be the primary avoidance mechanism for majority of the Starlink passages falling outside a given boresight cutoff, especially if a  more stringent limit ($<0.5^{\circ}$) is required. However, follow-up experiments with higher time resolution will be required to fully assess the effects of beam interactions between the telescope and the fast-moving satellites on the measured signal.

This work has provided critical data for both the NRAO and SpaceX teams to refine the design parameters of the telescope boresight-avoidance method, such as a boresight degree separation threshold and a satellite's downlink disabling timescale. In particular, feedback for NRAO's development of an autonomous self-reporting system, the Operational Data Sharing (ODS)\footnote{ODS data format \url{https://obs.vla.nrao.edu/ods/}} system, which will provide near real-time telescope operational data (e.g., pointing coordinate, observing band, observation mode and duration) to satellite operators to implement similar telescope avoidance schemes in their network algorithms. NRAO and SpaceX are planning real-time tests of an early version of this system in Summer 2024. Preliminary tests for the ODS system will be based on similar procedures for the coordinated experiments reported here. Importantly, NRAO's cooperative experimentation with SpaceX demonstrates the value of, and a pathway for, continuing cooperative efforts between passive (e.g., radio astronomy observatories) and active (e.g., satellite operators) spectrum users in order to achieve their respective goals.

The telescope boresight-avoidance method being developed by NRAO and SpaceX is a novel way to ensure the coexistence of radio astronomy and commercial satellite operators in a way that mutually benefits the mission of both groups. The initial results from this work suggest that these avoidance methods, when properly implemented and tested, can simultaneously increase the range of communication services of a satellite operator while expanding the frequency bands on which a radio astronomy telescope can observe without harmful interference from the satellite constellation. As boresight avoidance can in theory be implemented across the radio spectrum, there is good reason to think that a telescope that implements the method could observe on a much larger set of frequency bands than may be protected by existing regulations, while the satellite operator could simultaneously provide communication services in and around the telescope using other frequency bands that are not being observed by the telescope. This would mark a fundamental shift in thinking about the coexistence of radio astronomy and satellite services, where radio frequency regulations have traditionally imposed a strict separation of the two services through the hard allocation of certain frequency bands to one service at the exclusion of the other. The authors of the paper invite radio astronomers, satellite operators, and regulatory agencies to consider the implication of the telescope boresight avoidance method on radio frequency regulations and ongoing efforts to coordinate passive space research services and active satellite services.

\begin{acknowledgments}
The National Radio Astronomy Observatory is a facility of the National Science Foundation operated under cooperative agreement by Associated Universities, Inc. This work is supported by the National Science Foundation SII-NRDZ (AST-2232159) and SWIFT-SAT (AST-2332422) grants. The authors acknowledge the contributions of many individuals at both NRAO and SpaceX who have made these experiments possible. At NRAO and GBO: James Robnett, Fred Schwab, Sheldon Wasik, Daniel Bautista; and at SpaceX: Doug Knox, Michael Nicolls, David Goldman, David Partridge, Joe McMichael, Tony Liang, and Mihai Albulet.
\end{acknowledgments}

\bibliography{reference}{}
\bibliographystyle{aasjournal}



\end{document}